\documentclass[regular]{jfm}

\usepackage{graphicx}
\usepackage{newtxtext}
\usepackage{newtxmath}
\usepackage{natbib}
\usepackage{hyperref}
\usepackage{color}
\hypersetup{
    colorlinks = true,
    urlcolor   = blue,
    citecolor  = blue,
    linkcolor  = blue
}

\shorttitle{Inertial Particles in Turbulent Taylor-Couette Flow }
\shortauthor{H. Jiang and orthers}
\title{Spatial Distribution of Inertial Particles in Turbulent Taylor-Couette Flow}

\author{
  Hao Jiang\aff{1},
  Zhiming Lu\aff{1,2},\corresp{\email{zmlu@shu.edu.cn}}
  Bofu Wang\aff{1},
  Xiaohui Meng\aff{1},
  Jie Shen\aff{3}
	\and Kai Leong Chong \aff{1,2} \corresp{\email{klchong@shu.edu.cn}}
}

\affiliation{\aff{1} Shanghai Key Laboratory of Mechanics in Energy Engineering, Shanghai Institute of Applied Mathematics and Mechanics, School of Mechanics and Engineering Science, Shanghai University, Shanghai, 200072, China
\aff{2} Shanghai Institute of Aircraft Mechanics and Control, Zhangwu Road, Shanghai, 200092, China
\aff{3} Guangdong Provincial Key Laboratory of Turbulence Research and Applications, Center for Complex Flows and Soft Matter Research, and Department of Mechanics and Aerospace Engineering, Southern University of Science and Technology, Shenzhen 518055, China}

\begin{document}

\maketitle

\begin{abstract}
This study investigates the spatial distribution of inertial particles in turbulent Taylor-Couette flow. Direct numerical simulations are performed using a one-way coupled Eulerian-Lagrangian approach, with a fixed inner wall Reynolds number of 2500 for the carrier flow, while the particle Stokes number varies from 0.034 to 1 for the dispersed phase. We first examine the issue of preferential concentration of particles near the outer wall region. Employing two-dimensional (2D) Vorono{\"{i}} analysis, we observe a pronounced particle clustering with increasing $St$, particularly evident in regions of low fluid velocity. Additionally, we investigate the concentration balance equation, inspired by the work of \citet{johnson2020turbophoresis}, to examine particle radial distribution. We discern the predominant sources of influence, namely biased sampling, turbophoresis, and centrifugal effects. Across all cases, centrifugal force emerges as the primary driver, causing particle migration towards the outer wall. Biased sampling predominantly affects smaller inertial particles, driving them towards the inner wall due to sampling within Taylor rolls with inward radial velocity. Conversely, turbophoresis primarily impacts larger inertial particles, inducing migration towards both walls where turbulent intensity is weaker compared to the bulk. With the revealed physics, our work provides a basis for predicting and controlling particle movement and distribution in industrial applications.

\end{abstract}


\section{Introduction}\label{Introduction}
Particle-laden turbulent flows are prevalent in both natural phenomena and industrial applications, such as sandstorms \citep{di2018aerodynamic,zhang2020reconstructing}, combustors \citep{apte2003large}, and fluidized beds \citep{rokkam2010computational,kolehmainen2016hybrid}. These flows are inherently complicated due to the existence of turbulent fluctuation and also the boundary layer in shaping the intricate movement of particles. Despite its fundamental significance in multiphase flows, understanding the motion and distribution of particles within these systems is also important to efficiently manipulate the momentum or heat transports in industrial processes.

Previous studies investigating the interaction between turbulent flows and particles have often focused on homogeneous isotropic turbulence (HIT) as an idealized system. In HIT, it has been observed that inertial particles exhibit non-uniform spatial distribution characterized by clusters and voids, a phenomenon known as preferential concentration \citep{squires1991preferential,saw2008inertial,monchaux2012analyzing}. Through calculating the divergence of particle velocity, \citet{maxey1987gravitational} found that particles tend to accumulate in regions of high strain rate and low vorticity in the small Stokes number limit ($St \ll 1$). At first, this phenomenon was explained by the centrifugal effect caused by vortice in the turbulent flow \citep{abrahamson1975collision,reade2000effect}. As the particle inertia increases, the dominant mechanism of preferential concentration becomes the sling effect \citep{falkovich2002acceleration} and sweep-sticky mechanism \citep{goto2008sweep}.

Turbulent flows encountered in industrial settings are often bounded by solid walls. In such wall-bounded flows, the presence of no-slip and no-penetration boundary conditions leads to a sharp decrease in turbulence intensity near the wall. Consequently, particles disperse more rapidly in regions of higher turbulence intensity \citep{caporaloni1975transfer,reeks1983transport}, resulting in a tendency for particles to accumulate in the viscous sublayer at higher concentrations compared with other regions \citep{marchioli2002mechanisms,bernardini2014reynolds}. This phenomenon is known as turbophoresis, where couples of studies have indeed demonstrated the high particle concentration in the near-wall regions compared to the bulk \citep{marchioli2002mechanisms,sardina2012wall}. It has been observed that particle accumulation at the wall becomes pronounced when the particle relaxation time matches the local turbulence timescale. Additionally, numerical simulations have shown that particles moving away from the wall are associated with the ejection events \citep{vinkovic2011direct}. \citet{johnson2020turbophoresis} have recently examined the effects of biased sampling and turbophoresis on particle concentration profiles in turbulent channel flows, applying their model to wall-modeled large-eddy simulations.

In certain natural and industrial environments, additional forces come into play, further complicating the dynamical system. In systems subject to rotation, such as Taylor vortex reactors, centrifugal forces significantly alter the spatial distribution of particles by driving them away from the center of rotation. To investigate the combined effects of shear turbulence and rotation, the Taylor-Couette (TC) flow \citep{couette1890etudes,taylor1923viii,grossmann2016high} serves as a paradigmatic system. The TC flow consists of fluid confined in the gap between two rotating cylinders. The dimensionless parameters regarding the geometry can be described by the radius ratio $\eta = r_i/r_o$ and the aspect ratio $\varGamma = L/d$, where the $r_i$ ($r_o$) is the inner (outer) cylinder radius, $L$ and $d$ are the axial length and the gap between two cylinders, respectively. As the rotation of the inner and outer cylinders drive the flow, the two governing dimensionless parameters are the two Reynolds numbers \citep{grossmann2016high}, namely
\begin{gather}
  R e_{i,o}=\frac{{r_{i,o}} \omega_{i,o} d}{\nu}, \label{eq:1.1}
\end{gather}
where $Re_i$ and $Re_o$ are the Reynolds numbers of the inner and outer cylinders, respectively, $\omega_i$ and $\omega_o$ are the angular velocities of inner and outer cylinders, respectively, $\nu$ is the kinematic viscosity of fluid. When the outer cylinder is fixed, the relevant control paramter is solely given by the inner cylinder Reynolds number $Re_i = r_i \omega_i d / \nu$. Althernatively, one characterize the driving of the flow by Taylor number, which is 
\begin{gather}
  T a=\frac{(1+\eta)^4}{64 \eta^2} \frac{\left(r_{o}-r_{i}\right)^2\left(r_{i}+r_{o}\right)^2\left(\omega_{i}-\omega_{o}\right)^2}{\nu^2}, \label{eq:1.2}
\end{gather}
and regarding the rotation ratio, one use the inverse Rossby number $Ro^{-1}$.

The pioneering work of flow structure in TC flow is done by \citet{taylor1923viii}, who revealed the existence of Taylor vortex. Since then, four distinct flow pattens has been found at different control parameters, namely circular Couette flow (CCF), Taylor vortex flow (TVF), wavy vortex flow (WVF) and turbulent TC flow. Due to the rich variety of flow characteristics in TC flow, this system has gained significant attention. The focuses of the studies are the flow instabilities \citep{taylor1923viii,rudiger2018stability}, ultimate regimes \citep{froitzheim2019statistics,Hamede_2023}, property of inner and outer boundary layers \citep{brauckmann2017marginally}, pattern formation \citep{koschmieder1993benard}. There are also studies focusing the multiphase TC flow, where the important issue is the drag modulation led by dispersed phases, such as bubbles \citep{van2005drag,spandan2016drag,spandan2018physical}, droplets \citep{spandan2016deformation} and finite-sized particles \citep{wang_yi_jiang_sun_2022}. In the meantime, there are also several studies on examining the mixing problems in chemical processing applications \citep{schrimpf2021taylor} and biomedical field \citep{curran2005oxygen}.

TC flow is an ideal system for investigating dispersed multiphase flows with the interplay of shear turbulence and rotation. Studies on dispersed multiphase TC flow should date back to the 1920s \citep{taylor1923motion}. Sebsequently, many studies found the drifting of neutrally buoyant particle from high to low shear region in TC flows \citep{tetlow1998particle,fang2002flow,qiao2015particle,li2020eulerian,kang2021flow}. A comprehensive experiment conducted by \citet{majji2018inertial} observed the inertial migration of neutral particles in CCF, TVF and WVF with Reynolds numbers ranging from 83 to 151.4. They report the effect of the neutral particle inertial migration on the transition of TC flows \citep{baroudi2020effect}. \citet{li2020eulerian} adopted the Eulerian-Lagrangian simulation to examine the point particle equilibrium position in circular Couette flow with Reynolds numbers ranging from 60 to 90. \citet{dash2020particle} considered cases with larger paramter range with the Reynolds number extended to $O(10^3)$. They found the distinct role played by the particles where instead of simply destabilizing the flow, neutral particles may also decrease the growth of flow instabilities.


However, existing studies predominantly focus on particle densities close to that of the carrier phase, also often neglecting turbulent flow regimes in multiphase TC flow. The question of how particle inertia play a role in shaping their dispersion behaviour and distribution in multiphase TC flow remains poorly addressed. Therefore, the primary objective of this study is to explore the influence of particle inertia on the spatial distribution of particles in turbulent TC flow.

In the present study, we analyze the effect of particle inertia on particle preferential concentration. We further examine how varying densities of particles influence the radial distribution of particles. Our objective is to uncover the mechanisms governing particle radial concentration profiles at various $St$. The remainder of the paper is organized as follows. The governing equations and problem set-up are introduced in \S \ref{Numerical Setups}. Then, the results are shown in \S \ref{Result}. Concluding remarks are given in \S \ref{Conclusion}.


\section{Governing equations and problem set-up}\label{Numerical Setups}

\begin{figure*}
  \centering
  \includegraphics[width=0.35\linewidth]{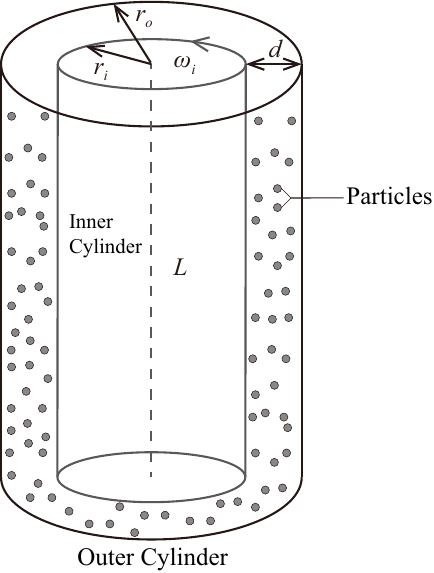}
  \caption{Sketch of the set-up of particle-laden Taylor-Couette flow. }
  \label{fig:1}
\end{figure*}

In the present investigation, we carry out direct numerical simulations (DNS) of there-dimensional (3D) particle-laden turbulent Taylor-Couette (TC) flow as shown in figure~\ref{fig:1}. 

\subsection{Carrier phase}

For the carrier phase, the governing equations are non-dimensionalized using the inner cylinder rotation speed, $\omega_i r_i$, and gap of cylinders, $d$, which are given by:
\begin{gather}
  \frac{\partial \pmb{u}}{\partial t} + (\pmb{u} \cdot \nabla) \pmb{u}=-\nabla p+\frac{1}{Re_i} \nabla^{2} \pmb{u}, \label{eq:2.1}\\
  \nabla \cdot \pmb{u}=0. \label{eq:2.2}
\end{gather}
In the equation (\ref{eq:2.1}), $Re_i = d \omega_i  r_i/\nu$ is the Reynolds number defined by the inner cylinder rotation.

DNS of the carrier phase is performed using a second-order accurate finite-difference method in cylindrical coordinates $( \pmb{e}_r, \pmb{e}_{\theta}, \pmb{e}_z)$ \citep{verzicco1996finite,ostilla2013optimal}  with a uniform grid spacing in the azimuthal and axial directions and a non-uniform grid spacing by a clipped Chebychev type clustering method in the radial direction. The computational domain is $r_i \le r \le r_o$, $0 \le \theta \le 2\pi$ and $0 \le z \le L_z$, where $r_i$ and $r_o$ are the inner and outer cylinder radius, respectively, and $L_z$ is the axial length of the domain. The no-slip boundary condition is applied at the inner and outer cylinder walls, and the periodic boundary condition is applied in the azimuthal and axial direction.

\begin{table}
  \setlength{\tabcolsep}{0.2cm}
  \begin{center}
\def~{\hphantom{0}}
  \begin{tabular}{lcccccccc}
      $Re_{i}$ & $N_{\theta} \times N_r \times N_z$ & $\varGamma$ & $\eta$ &$Nu_{\omega}$ &$Nu_{\omega}$ &$\epsilon$(\%)\\[3pt]
             &    &   & & & (present study)& compared with \citet{ostilla2013optimal}\\ 
       160     & 128$\times$ 64 $\times$ 64         &  2$\pi$          & 0.714      &1.878         &1.869 &0.48 \\
       2500    &  384 $\times$ 192 $\times$ 192      & 2$\pi$      & 0.714  &6.443         &6.421 &0.34 \\
  \end{tabular}
  \caption{Validation of Taylor-Couette simulations at $Re_i=160,2500$. $Re_i$ is the Reynolds number of the inner cylinder, $N_{\theta} \times N_r \times N_z$ the grid resolution, $\varGamma$ the aspect ratio, $\eta$ the radius ratio, $Nu_{\omega}$ the normalized angular velocity flux, $\epsilon$ is the relative error of $Nu_\omega$ comparing with \citet{ostilla2013optimal}.}
  \label{tab:flowValidation}
  \end{center}
\end{table}

The numerical approach has been adopted in many of our previous studies to simulate the problem of single phase and multiphase turbulent flows under Cartesian geometry \citep{zhao2022modulation,zhao2022suppression,guo2023flow,meng2024simulation}.
In order to further validate our numerical code within cylindrical geometry, we perform simulation of TC flow with the same grid resolutions and control parameters as in \citet{ostilla2013optimal}, and compare our results of $Nu_\omega$ in table~\ref{tab:flowValidation}. Here the $Nu_\omega$ is defined as,
\begin{gather}
N u_\omega=\frac{J}{J_{l a m}}, \quad \text { with } J=r^3\left( \left\langle u_r \omega\right\rangle_{A, t}-v \partial_r\langle\omega\rangle_{A, t}\right),
\end{gather}
where $\langle \cdot \rangle_{A, t}$ denotes the azimuthal, axial and time average, and $J_{lam}$ is the angular velocity flux for laminar flow. The results are in good agreement with the results of \citet{ostilla2013optimal}.

\subsection{Lagrangian particle tracking}
In this study, we consider the inertial particles using Lagrangian point particle approach \citep{maxey1983equation,maxey1987gravitational,Gatignol_jmta_1983,Tsai_taml_2022}. 
Particles are released into the flow when the turbulent flow was fully developed. Low volume fraction of particles has been considered, and thus the feedback of particles to the flow field and the particle-particle collisions has been neglected \citep{elghobashi1991particle,elghobashi1994predicting,elgobashi2006updated,balachandar2010turbulent}. Small diameter $d_p$ and large density ratio $\rho^* = \rho_p/\rho_f$ are considered in this study. The Lagrangian particle tracking considers Stokes drag in the governing equation of particle motion. The dimensionless equations of particle translational motion in cylindrical coordinates read as:
\begin{gather}
  \frac{\mathrm{d} v_r}{\mathrm{d} t}=\frac{C_D}{St}\left(u_r-v_r\right)+\frac{v_\theta^2}{r}, \label{eq:2.4}\\
  \frac{\mathrm{d} v_\theta}{\mathrm{d} t}=\frac{C_D}{St}\left(u_\theta-v_\theta\right)-\frac{v_\theta v_{\mathrm{r}}}{r}, \label{eq:2.5}\\
  \frac{\mathrm{d} v_z}{\mathrm{d} t}=\frac{C_D}{St}\left(u_z-v_z\right). \label{eq:2.6}
\end{gather}
where $C_D = 1+ 0.15Re_p^{0.687}$ is the drag coefficient, $Re_p = d_p^* |\pmb{u} -  \pmb{v}|Re_i$ is the particle Reynolds number, where $\pmb{v}$ and $\pmb{u}$ are the particle and fluid dimensionless velocity, respectively. The particle diameter $d_p^*$ is normalized by the gap width $d$. $v_r$, $v_\theta$ and $v_z$ are the particle velocity in the radial, azimuthal and axial directions, respectively. The particle Stokes number is defined as $St = \rho^* d_p^{\ast2} Re_i/18$, where $\rho^* = \rho_p/\rho_f$ is the density ratio between the particle and fluid. The particle Stokes number is a dimensionless parameter that describes the ratio of the particle relaxation time to the characteristic time scale of the flow. The particle Stokes number and Reynolds number are two important parameters that describe the particle motion in the flow field. In this study, the particle Stokes number is fixed at $St=0.034,0.17,0.34,0.68,1$ as shown in table~\ref{tab:particle setup}. The particle diameter is fixed at $d_p^*=0.005$ and volume fraction is fixed at $5.5 \times 10^{-5}$, and at this small volume fraction, one-way coupling is sufficient for describing the motion of the particles. The particle density ratio increases from 10 to 288 to study the distribution of particles with different inertia.

\begin{table}
  \setlength{\tabcolsep}{0.3cm}
  \begin{center}
\def~{\hphantom{0}}
  \begin{tabular}{lcccccc}
    
      Case  & $d_p^*$ & $\phi_v$  & $\rho^*$ & $St$ \\[3pt]
       1    & 0.005   & $5.5 \times 10^{-5}$ & 10       & 0.034     \\
       2    & 0.005   & $5.5 \times 10^{-5}$ & 50       & 0.17      \\
       3    & 0.005   & $5.5 \times 10^{-5}$ & 100      & 0.34      \\
       4    & 0.005   & $5.5 \times 10^{-5}$ & 200      & 0.68    \\
       5    & 0.005   & $5.5 \times 10^{-5}$ & 288      & 1     \\
  \end{tabular}
  \caption{Details of inertial particle in the numerical simulations. The $d_p^* = d_p / d$ is the particle diameter normalized by the gap width, $\phi_v$ is the volume fraction, $\rho^* = \rho_p/\rho_f$ the density ratio between the particle and fluid, $St=\rho^\ast d_p^{\ast2}Re_i/18$ the particle Stokes number.}
  \label{tab:particle setup}
  \end{center}
\end{table}

\section{Result and discussion} \label{Result}

\begin{figure*}
  \centering
  \includegraphics[width=1\linewidth]{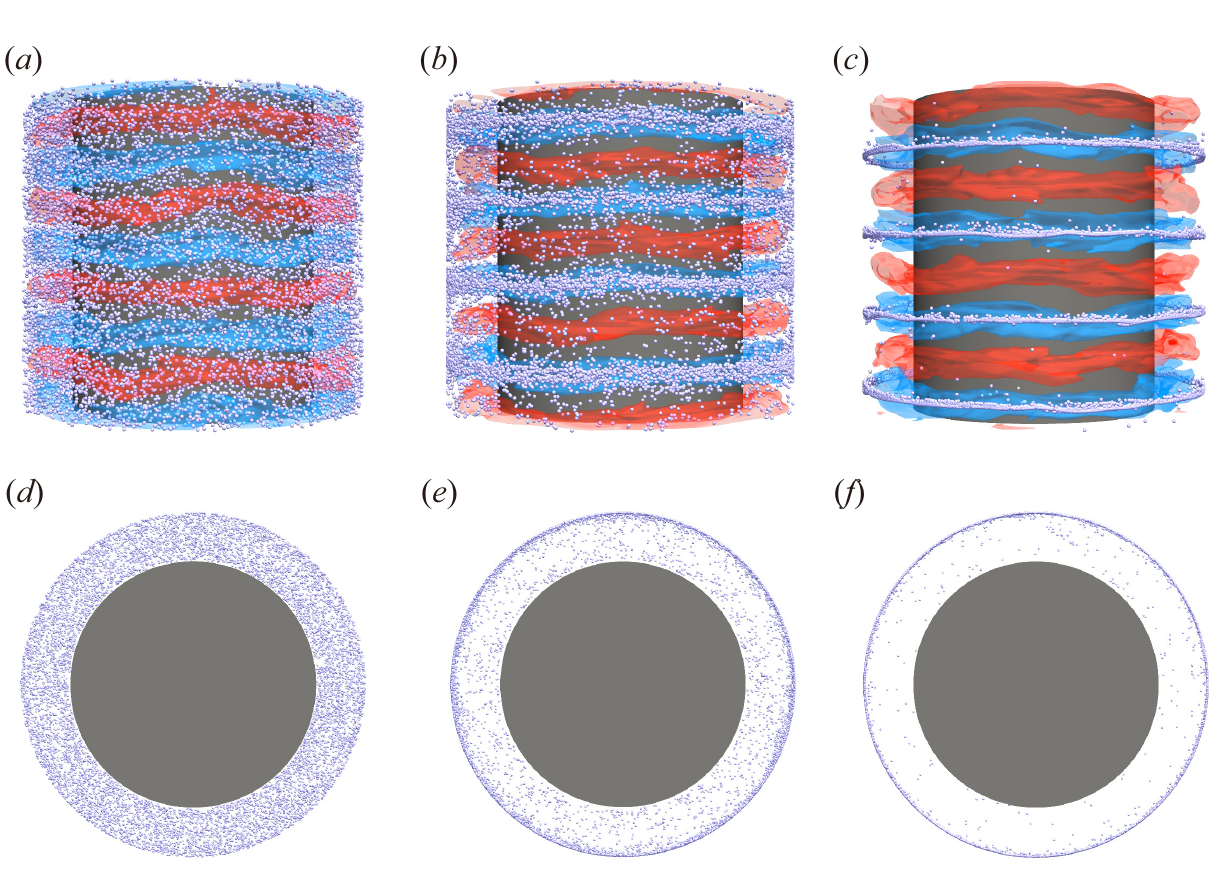}
  \caption{Three-dimensional visualization ($a$-$c$) and top view ($d$-$f$) instantaneous snapshots of particle distribution for the ($a$,$c$) $St=0.034$, ($b$,$e$) $St=0.34$, ($c$,$f$) $St=1$; The iso-surface of radial velocity has also been drawn in the 3D visualization ($a$-$c$) where the reddish surface represents the outward velocity and the bluish one represents the inward velocity, particularly at the dimensionless radial velocity $-0.1$ and $0.1$. }
  \label{fig:2}
\end{figure*}

\subsection{Particles preferential concentration} \label{sub:3.2}

To study how the particle inertia influences its distribution in Taylor-Couette flow, we first present the 3D visualization of the instantaneous particle distributions with varying Stokes numbers, as illustrated in figure~\ref{fig:2}. Different patterns of particle distribution emerge depending on $St$. When the particle inertia is minimal ($St=0.034$), the particles are dispersed throughout the entire domain. With an increase in particle inertia (i.e. with higher $St$), particles tend to cluster, forming distinct parallel strip-like structures in the axial direction. For example, four clearly defined strips become evident at $St=1$ as seen in figure~\ref{fig:2}($c$). We also draw the iso-surface of in radial velocity in the 3D visualization which shows the footprint of Taylor rolls---a hallmark feature of Taylor-Couette flow. One sees that those strip-like particle clusters coincide with the inward velocity region of Taylor rolls. From the top-view of particle distribution shown in figure~\ref{fig:2}($d$-$f$), it is observed the the stripped particle cluster located near the outer walls as the particle inertia becomes more pronounced with the increasing $St$. 

\begin{figure*} 
  \centering
  \includegraphics[width=0.8\linewidth]{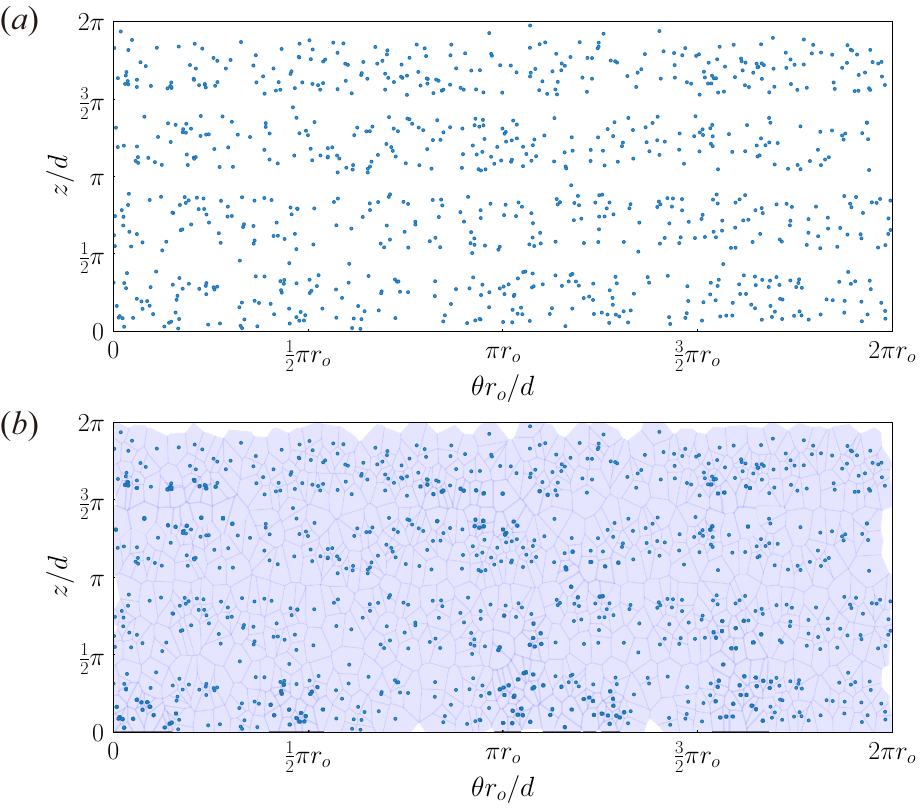}
  \caption{($a$) Instantaneous snapshot of particle distribution at near outer wall region for the $St = 0.034$, ($b$) the 2D Vorono{\"{i}} diagrams corresponding to the ($a$).}
  \label{fig:3}
\end{figure*}

In numerous industrial applications, the accumulation of particles near the wall region often leads to collisions or adhersions between particles, making this phenomenon a key aspect to examine. To characterize the particle distribution in the vicinity of the outer wall, we utilize a two-dimensional Vorono{\"{i}}  diagrams analysis to quantify the preferential concentration of particles. The Vorono{\"{i}}  diagram is a spatial tessellation in which each Vorono{\"{i}}  cell is defined at the particle location based on the distance to adjacent particles. Consequently, in regions where particles cluster, the area of the Vorono{\"{i}} cells is smaller compared to those in adjacent regions. For the analysis of particle distribution in the near outer wall region, we consider particles within the range $r_0-1.5d_p^* \le r \le r_0-0.5d_p^*$. As an example, figure~\ref{fig:3} illustrates the 2D Vorono{\"{i}} diagrams depicting the instantaneous distribution of particles near the outer wall for $St = 0.034$.

\begin{figure*} 
  \centering
  \includegraphics[width=1\linewidth]{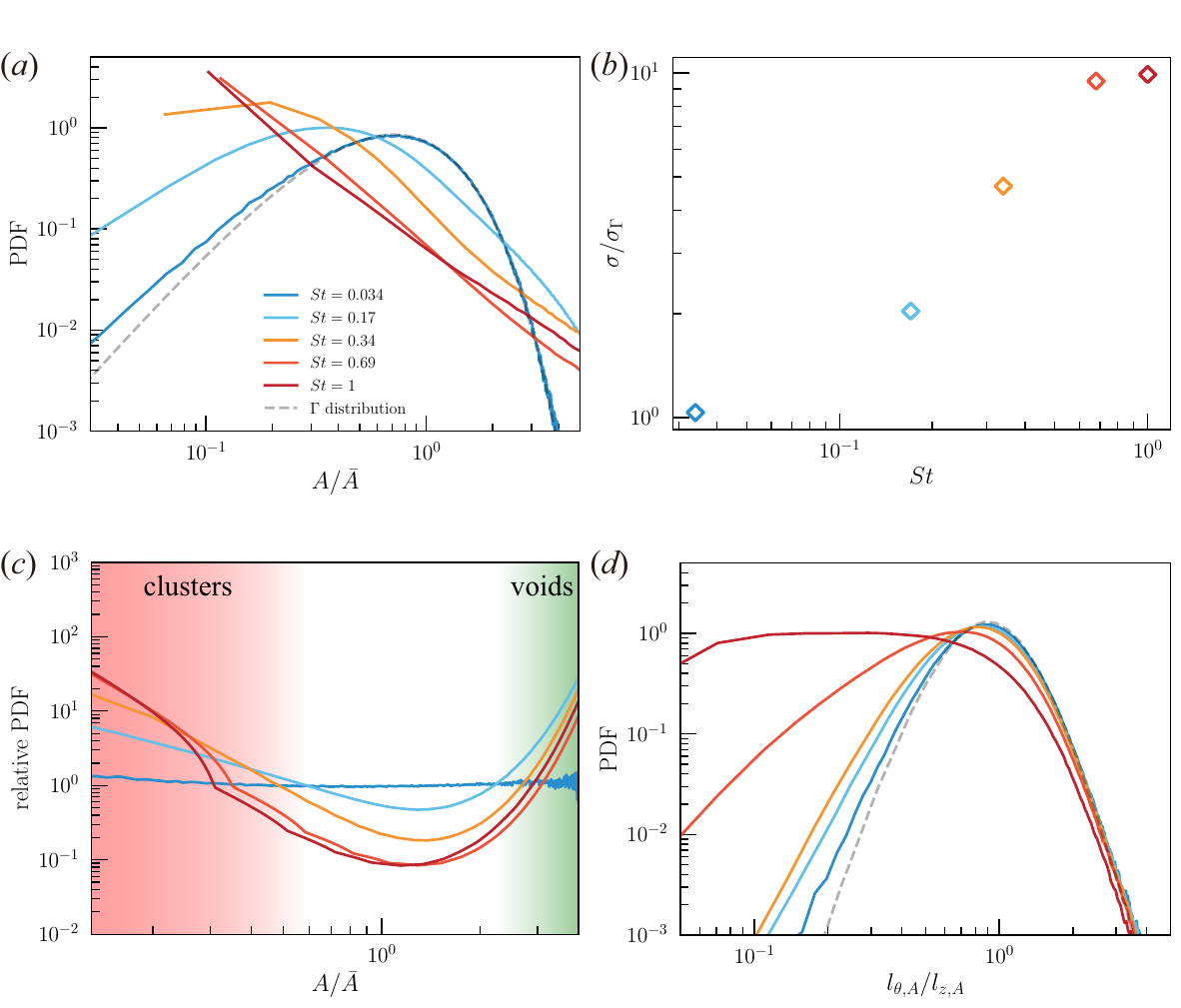}
  \caption{Statistics of 2D Vorono{\"{i}} diagrams for particles near the outer wall at various Stokes numbers: ($a$) the normalized Vorono{\"{i}} area PDFs, where the grey dashed curve shows the $\Gamma -$distribution for 2D random distribution; ($b$) the normalized standard deviation $\sigma / \sigma_{\Gamma}$ of Vorono{\"{i}} area distribution; ($c$) the relative PDF defined by the ratio of the PDFs to the  2D $\Gamma -$distribution counterpart. Here, the reddish (greenish) area denotes the regime having cluster (void) of particles; ($d$) the PDFs of Vorono{\"{i}} cells aspect ratio, defined by ratio of azimuthal length to the axial length, where the grey dashed curve shows the case for 2D random distribution.}
  \label{fig:4}
\end{figure*}

As the area of the Vorono{\"{i}} cells can represent particle clustering, we examine the Probability Density Functions (PDFs) of Vorono{\"{i}} cell areas to quantify the preferential concentration of particles. The PDFs of particle Vorono{\"{i}} area in the near outer wall region are presented in figure~\ref{fig:4}($a$), with the solid lines representing the PDFs of simulation data. The dashed line corresponds to the PDF of Vorono{\"{i}} cell areas (normalized by the mean area) for randomly distributed particles, as described by a $\Gamma$-distribution \citep{ferenc2007size}. 

The $\Gamma$-distribution of two-dimensional Vorono{\"{i}} area is given as,
\begin{gather}
  f(A^+)=\frac{343}{15} \sqrt{\frac{7}{2 \pi}} (A^{+})^{5 / 2} \exp \left(-\frac{7}{2} A^{+}\right). \label{eq:3.8}
\end{gather}
Here, $A^+ = A/\bar{A}$ represents the Vorono{\"{i}} cell area normalized by the mean value. At $St=0.034$, the PDF of the Vorono{\"{i}} cell area closely aligns with the $\Gamma$-distribution, with only slight deviations in the regime where  $A/\overline{A}<0.6$. However, as the particle $St$ number increases, the deviation from  $\Gamma$-distribution becomes more pronounced. Moreover, the PDF values are not only larger than the $\Gamma$-distribution in the small area regime but also in the large area regime. Our results suggest that the preferential concentration of particles becomes increasingly significant with an increase in $St$. Next, we investigate the variation of the standard deviation of the Vorono{\"{i}} cell area normalized by the standard deviation of the $\Gamma$-distribution \citep{ferenc2007size}, with respect to the particle Stokes number ($St$), as illustrated in figure \ref{fig:4}($b$). By comparing $\sigma/\sigma_{\Gamma}$ across different $St$ values, we observe a clear trend of increasing particle preferential concentration with higher Stokes numbers.

To closely inspect the particle preferential distribution, we examine the ratio of respective PDFs to the $\Gamma$-distribution PDF counterpart, namely the relative PDF \citep{monchaux2010preferential} in figure~\ref{fig:4}($c$). For each relative PDF, there are two intersections point with the value $1$. According to \cite{monchaux2010preferential}, these intersections present the formation of clusters (red) and voids (green). With the increase in $St$, particle clusters become more pronounced, which is consistent with the results shown in figure \ref{fig:4}($a$) and \ref{fig:4}($b$).

\begin{figure*} 
  \centering
  \includegraphics[width=1\linewidth]{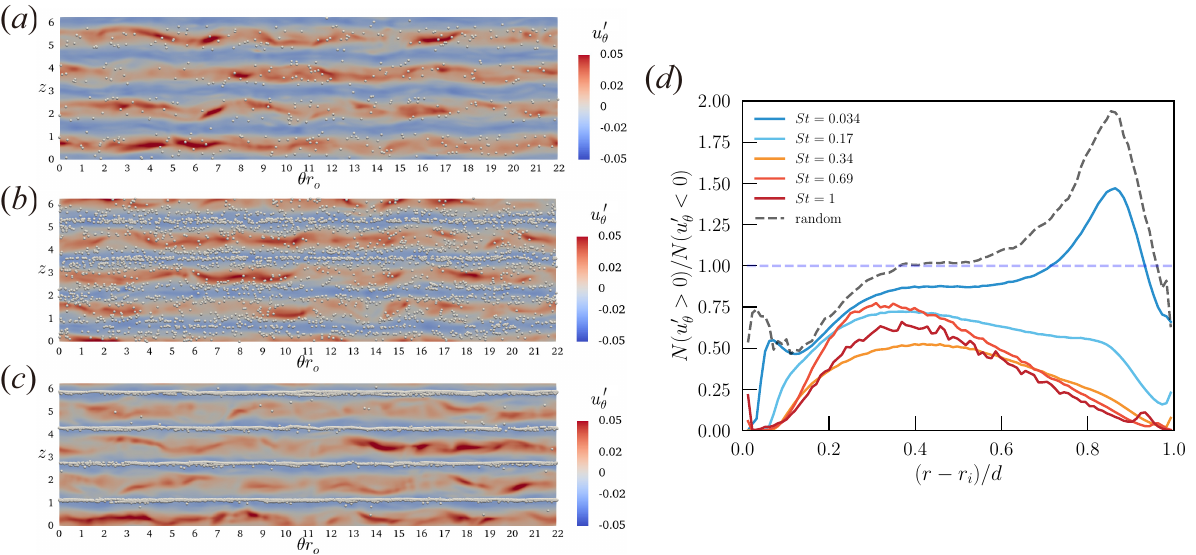}
  \caption{($a$)-($c$) Snapshots depicting instantaneous distributions of particles at different Stokes numbers in the near outer-wall ($\theta-z$) plane, with the color representing fluctuation of azimuthal velocity $u^{\prime}_{\theta}$ of the fluid. ($d$) Radial profiles of the particle numbers with $u^{\prime}_{\theta}>0$ to those with $u^{\prime}_{\theta}<0$. The gray dashed curve represents the result for a random distribution of particles.}
  \label{fig:5}
\end{figure*}

In addition, to further quantify the characteristics of the particle distribution in the outer wall, we examine the PDF of the aspect ratio of the Vorono{\"{i}} cell: $l_{\theta,A}/l_{z,A}$. $l_{\theta,A}$ is the maximum length in the azimuthal direction of a Vorono{\"{i}} cell and $l_{z,A}$ is the maximum length in the axial direction. In Figure \ref{fig:4}($d$), it is evident that the area enclosed by the probability density function (PDF) curve within the range $l_{\theta,A}/l_{z,A}$ is greater than that within the range of $l_{\theta,A}/l_{z,A}$. This indicates that particles prefer to cluster in the azimuthal direction, which can also be seen in figure~\ref{fig:5}($a$)-($c$). In addition, as $St$ increases, the left tail of the PDF gets higher and the peak shifts to the left. This suggests that the greater the inertia of the particles, the more pronounced the aggregation of particles on the outer wall and the formation of an azimuthal stripe structure.

Subsequently, we analyze the radial profile of the ratio of particle numbers with $u^{\prime}_{\theta}>0$ to those with $u^{\prime}_{\theta}<0$, as depicted in Figure \ref{fig:5}($d$). We can see from figure~\ref{fig:5}($d$) that for particles with $St>0.034$, the number of particles with $u^{\prime}_{\theta}>0$ is lower than that with $u^{\prime}_{\theta}<0$, implying that particles prefer to accumulate in low-speed streaks. For particles with $St=0.034$, their trajectory is very close to that of the fluid particle (i.e., the tracer particle) due to their small inertia, and their velocity characteristics are close to those of randomly distributed particles.

\subsection{Profile of particle radial concentration} \label{sub:3.1}


In wall-bounded particle-laden turbulent flow, the wall-normal particle distribution is also worthy to be studied. The presence of wall boundary affects the movement and distribution of particles along the wall-normal direction. To examine the underlying mechanism governing the radial distribution of particle concentration in particle-laden TC flow and the dominant forces, we employ a theoretical model inspired by \citet{johnson2020turbophoresis}, where their work considers the situation of plane Couette flow and disentangled the dominant effects into biased sampling and turbophoresis. 

Following the concept by \citet{johnson2020turbophoresis}, we derived the corresponding disentangled relation for the particle concentration in particle-laden TC flow with an additional centrifugal term:
\begin{gather}
  C^*\left(r\right)=\alpha \exp \left(\underbrace{\frac{1}{St} \int^{r} \frac{\left\langle u_{r} |\eta\right\rangle}{\left\langle v_r^2 |\eta\right\rangle} \mathrm{d} \eta}_{\text {biased sampling }}-\underbrace{\int^{r} \frac{\mathrm{d} \ln \left\langle v_{r}^{2} | \eta\right\rangle}{\mathrm{d} \eta} \mathrm{d} \eta}_{\text {turbophoresis }}+\underbrace{\int^{r} \frac{\left\langle \frac{v_{\theta}^2}{r} | \eta\right\rangle}{\left\langle v_{r}^{2} |\eta\right\rangle} \mathrm{d} \eta}_{\text {centrifugal effect }}\right),
\label{eq:3.1}
\end{gather}
which is derived from the radial direction single-particle position-velocity probability density function (PDF). Detail of the derivation can be find in Appendix~\ref{Appendix}. The numerical expression for the dimensionless concentration $C^*(r)$ is:
\begin{gather}
  C^*\left(r\right)= \frac{n_rV}{n_0V_{slice}}. 
\label{eq:3.2}
\end{gather}
Here, $n_r$ and $V_{slice}$ represent the particle numbers in each slice and slice volume. $n_0$ and $V$ are the total particle number and the volume of the entire domain. $u_r$ is the radial component of fluid velocity sensed by particles, and $v_r$ and $v_{\theta}$ are the radial and angular components of the particle velocity respectively. $\alpha$ is an integration constant. $\left\langle  \cdot  \right\rangle$ is the ensemble average done by averaging over all particles in the slice at corresponding radial position. There are three terms known as phoresis integrals in right-hand side of the equation (\ref{eq:3.1}). The first term is the effect of biased sampling on the particle concentration profile:
\begin{gather}
  I_b= \frac{1}{St}\int^{r} \frac{\left\langle u_{r} |\eta\right\rangle}{\left\langle v_r^2 |\eta\right\rangle} \mathrm{d} \eta. 
\label{eq:3.3}
\end{gather}
The second term represents the effect of turbophoresis on the particle concentration profiles: 
\begin{gather}
  I_t= - \int^{r} \frac{\mathrm{d} \ln \left\langle v_{r}^{2} | \eta\right\rangle}{\mathrm{d} \eta} \mathrm{d} \eta. 
\label{eq:3.4}
\end{gather}
The last term is given by
\begin{gather}
  I_c= \int^{r} \frac{\left\langle \frac{v_{\theta}^2}{r} | \eta\right\rangle}{\left\langle v_{r}^{2} |\eta\right\rangle} \mathrm{d} \eta, 
\label{eq:3.5}
\end{gather}
which quantifies the centrifugal effect on the particle radial distribution.

\begin{figure*}
  \centering
  \includegraphics[width=0.7\linewidth]{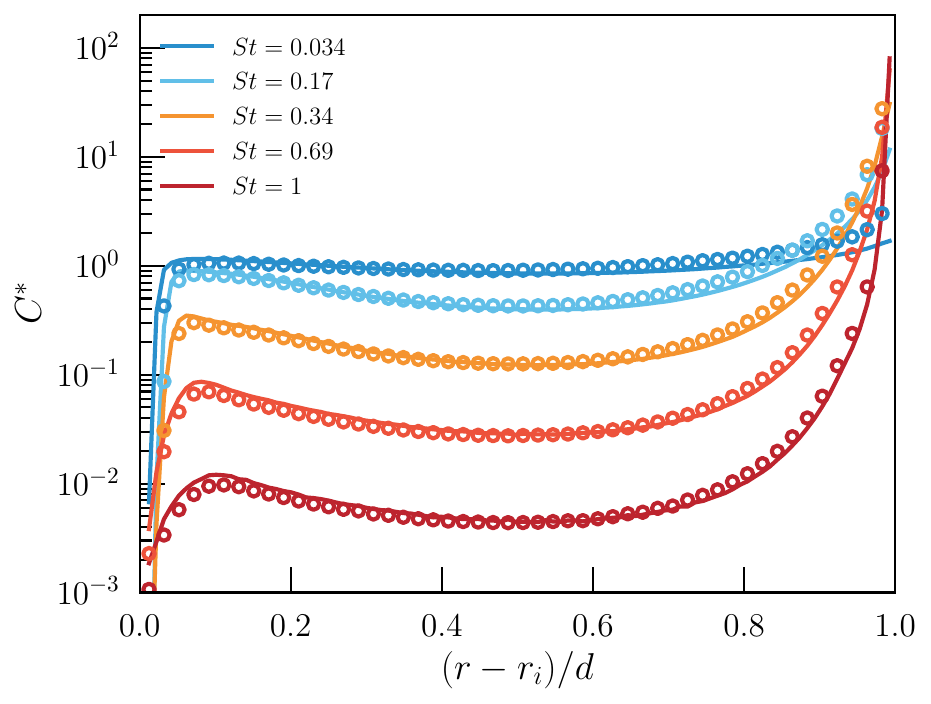}
  \caption{Radial profiles of particle concentration at various Stokes numbers. The curves are the results from DNS and the symbols represent the results based on the equation (\ref{eq:3.1}).}
  \label{fig:6}
\end{figure*}

Figure~\ref{fig:6} shows the particle concentration profiles with various Stokes numbers ($St$) at $Re_i = 2500$. The figure displays both the data obtained from simulation (solid curves) and the theoretical points (open symbols) according to the equation~(\ref{eq:3.1}). It shows that the equation can nicely describe the variation of the particle concentration. The concentration of particle decreases significantly near the inner wall and in the bulk with the increasing $St$, whereas there is substantial increase near the outer wall.

\begin{figure*}
  \centering
  \includegraphics[width=1\linewidth]{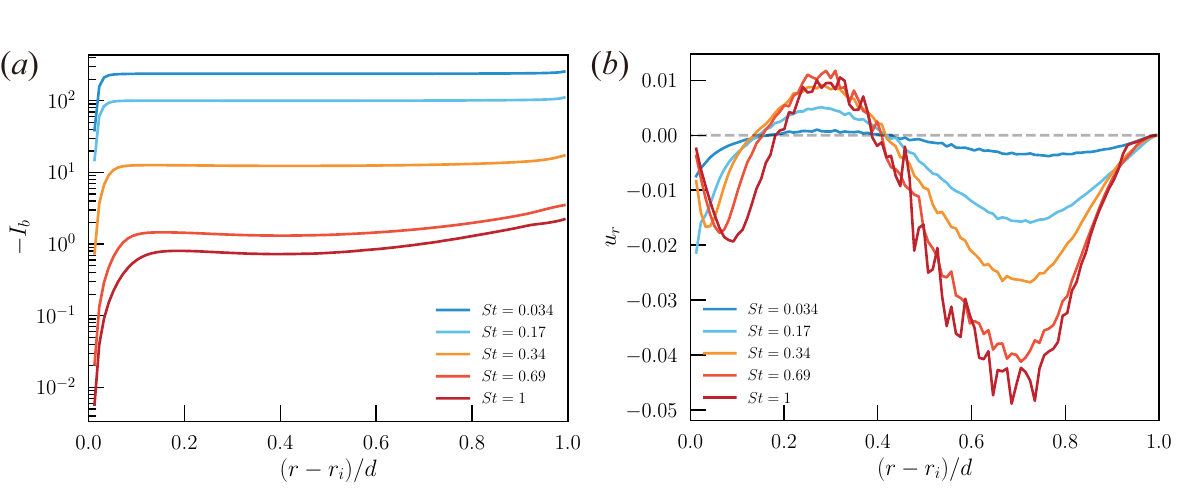}
  \caption{($a$) Radial profiles of negative of biased sampling integral (\ref{eq:3.3}); ($b$) Radial profiles of fluid radial velocity at particle location. }
  \label{fig:7}
\end{figure*}

\begin{figure*}
  \centering
  \includegraphics[width=1\linewidth]{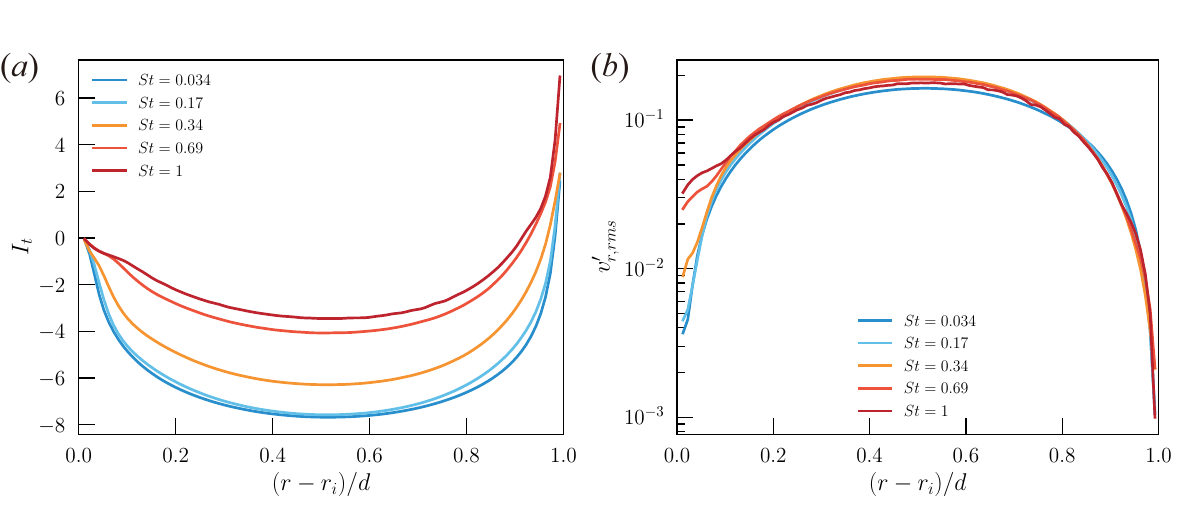}
  \caption{($a$) Radial profiles of turbophoresis integral (\ref{eq:3.4}); ($b$) Radilal profiles of root-mean-square radial velocity of particle. }
  \label{fig:8}
\end{figure*}

\begin{figure*}
  \centering
  \includegraphics[width=1\linewidth]{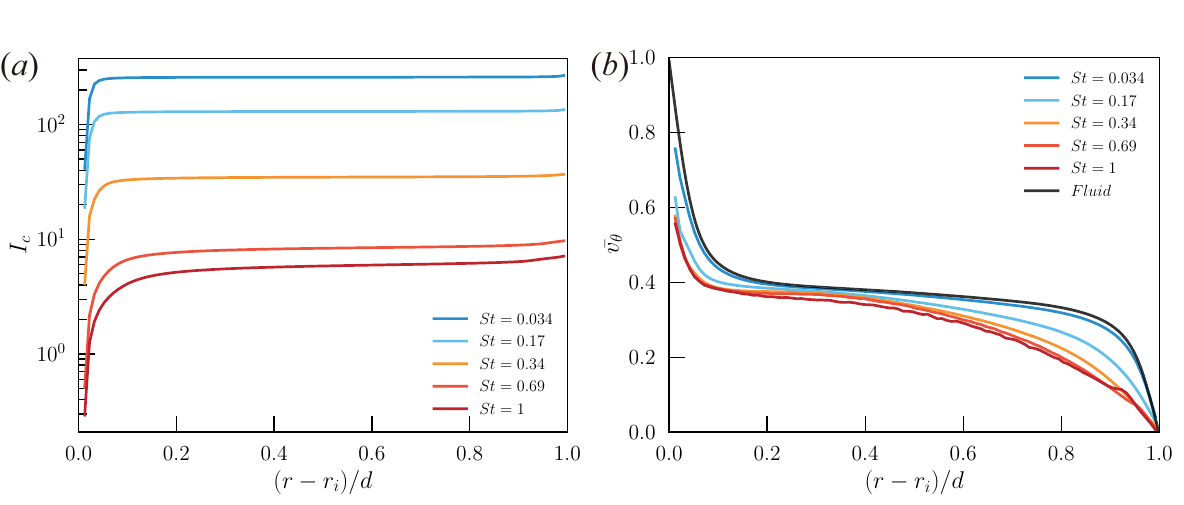}
  \caption{($a$) Radial profiles of centrifugal integral (\ref{eq:3.5}); ($b$) Radial profiles of azimuthal velocity of particle. As the reference, the black curve shows the radial profile of the fluid azimuthal velocity.}
  \label{fig:9}
\end{figure*}

The benefit on using equation~(\ref{eq:3.1}) is to disentangle the contributions from biased sampling, turbophoresis and centrifugal effect. In such way, one can examine how these three effects come into play in different regions of turbulent TC flow (near inner wall, outer wall and in the bulk). Thus, we examine the radial profiles of sampling bias $I_b$ (\ref{eq:3.3}), turbophoresis $I_T$ (\ref{eq:3.4}) and centrifugal effect $I_c$ (\ref{eq:3.5}) integrals, as shown in figure~\ref{fig:7}, \ref{fig:8}, and \ref{fig:9} respectively. 

The first phoresis integral accounts for the average drag force on the ensemble of particles at a given radial location. As depicted in Equation (\ref{eq:3.3}), it is proportional to $\left\langle u_{r} |\eta\right\rangle$, which represents the fluid radial velocity sampled at the particle position. For small $St$, a relatively random distribution of particles can be observed as shown in figure \ref{fig:2}($a$), resulting in a vanished drag term. Conversely, with sufficiently large $St$, where particle inertia becomes dominant, heavy particles tend to accumulate in regions with radially-inward velocity. It is notable that for plane Couette flow, this biased sampling also occurs for large $St$, causing heavy particles to gather in low-speed streaks in the ejecting region \citep{rashidi1990particle,eaton1994preferential,marchioli2002mechanisms}.

In Taylor-Couette flow, as heavy particles cluster around regions with inward radial velocity ($u_r<0$) more than those with outward velocity ($u_r>0$), it results in a negative value of the biased sampling term. In figure~\ref{fig:7}($a$), we plot $-I_b$ in logarithmic scale versus radial position, also including the averaged fluid velocity versus radial position for reference in figure~\ref{fig:7}($b$). Indeed, in the bulk, the particles experience progressively stronger inward radial velocity for larger $St$, causing $I_b$ to monotonically decrease in the radial direction. This indicates that the biased sampling provides a net force that pushes the particles towards the inner walls.

The second phoresis integral addresses the turbophoresis pseudo-force, accounting for the migration of particles to regions with smaller radial velocity variance. With the no-slip and non-penetration boundary conditions imposed at the inner and outer walls, the radial velocity variance vanishes at both walls, leading to a tendency for particle migration toward the walls. In figure \ref{fig:8}($a$), the turbophoresis integrals are plotted for different $St$, while the root-mean-square (rms) radial particle velocity ($v_{r,rms}^{\prime}$) has also been depicted in figure \ref{fig:8}($b$). Notably, due to the vanished velocity variance at the two walls, a minimum $I_t$ is indeed observed in the bulk. It is apparent that the velocity variance declines at a greater rate at the outer wall than at the inner wall, resulting in a larger $I_t$ at the outer wall compared to the inner wall. This indicates a greater tendency to attract particles to the outer wall relative to the inner wall.

Finally, the radial dependency of the centrifugal effect (\ref{eq:3.5}) can be observed in figure~\ref{fig:9}($a$), showing a monotonic increase of $I_c$ along the radial direction, irrespective of $St$. The rotation of the inner wall causes particles to gain angular velocity, as depicted in figure ~\ref{fig:9}($b$), illustrating the radial profiles of the azimuthal velocity of particles. Consequently, particles exhibit a tendency to migrate outward in the domain due to the centrifugal effect.

\begin{figure*}
  \centering
  \includegraphics[width=1\linewidth]{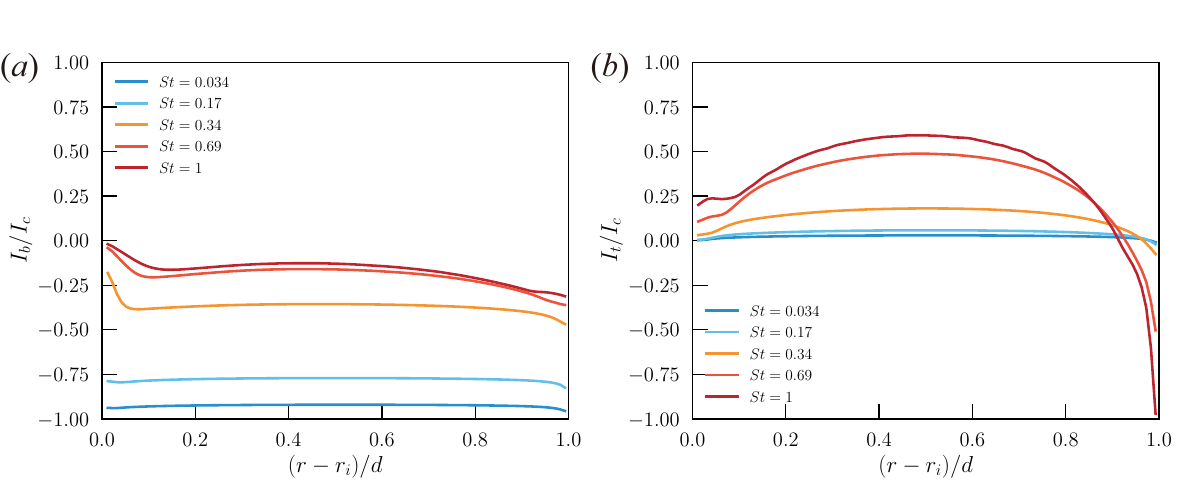}
  \caption{
  ($a$) Radial profiles of the ratio of biased sampling integral to centrifugal integral; ($b$) Radial profiles of the ratio of turbophoresis integral and centrifugal integral.}
  \label{fig:10}
\end{figure*}

We now comprehend the distinct roles of biased sampling, turbophoresis, and centrifugal effect in driving particles in TC flow: biased sampling leads to the migration of particles toward the inner wall, in contrast to the roles played by centrifugal and turbophoresis, where particles move toward the outer wall and both walls, respectively. To grasp the overall effect, we compare the relative dominance of the three mechanisms in particle concentration profiles, by examining the ratios of the intensities $I_b/I_c$ and $I_t/I_c$, as depicted in figure~\ref{fig:10}. From the figure, it is evident that the magnitude of both quantities is consistently less than one, signifying that the centrifugal effect holds the greatest strength at any radial position. For the smallest explored $St$ ($=0.034$), the strength of biased sampling is nearly equivalent to that of centrifugal, while the effect of turbophoresis is negligibly small. As a result, with two competing effects of similar strength, the particle concentration remains relatively uniform throughout the domain. However, the situation of relative dominance changes for heavy particles (large $St$). The relative strength of biased sampling progressively weakens with increasing $St$. Concurrently, turbophoresis becomes significant, with the magnitude of $I_t/I_c$ exceeding 60\% in the bulk and reaching almost 100\% near the outer wall. Indeed, a noticeable accumulation of heavy particles near the outer wall is observed due to the combined effect of centrifugal and turbophoresis, pushing heavy particles toward that wall.

\section{Concluding remarks and outlook}\label{Conclusion}

\begin{figure*} 
  \centering
  \includegraphics[width=1\linewidth]{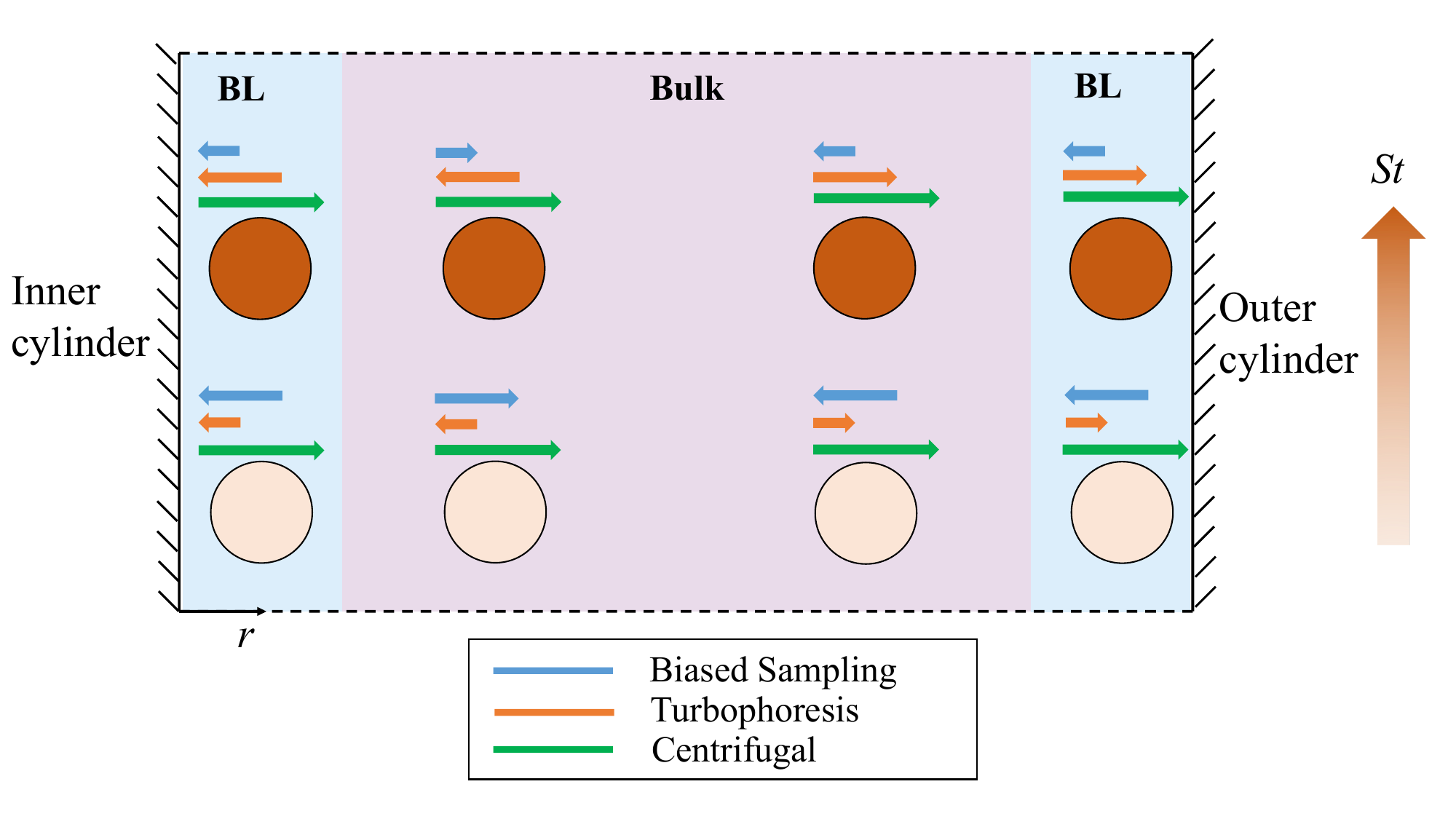}
  \caption{Schematic diagram showing the dorminant terms in shaping the radial distribution of inertial particles at various $St$ in multiphase Taylor-Couette flow. The bluish and pinkish zone present the region of boundary layers and bulk, respectively. Three arrows of distinct colors represent represent the relative strength of the biased sampling, turbophoresis and centrifugal effect. Here, the length of arrow present the intensity of the effect, and the directions of arrows represent the direction of particle migration led by the corresponding effect.}
  \label{fig:11}
\end{figure*}

To investigate the impact of particle inertial on particle distribution in wall-bounded turbulent flows, we conducted direct numerical simulations of particle-laden turbulent Taylor-Couette flow at $Re_i=2500$, introducing particles with various Stokes numbers. The simulations employed one-way coupled Lagrangian tracking of particles, with particle motions solely governed by the Stokes drag. Given the very low solid volume fraction considered, both two-way coupling and inter-particle collisions were disregarded. The focus of the study is on the spatial distribution of particles in TC flow. We have first examined the particle distribution near the outer cylindrical wall via 2D Vorono{\"{i}} diagram analysis given that particles tend to cluster around that wall for large enough $St$. Then, the approach ultilizing the conservation equation for particle concentration has been employed, which is originally proposed by \cite{johnson2020turbophoresis}. The benefit of this approach is to disentangle the effect caused by  biased sampling, turbophoresis, and centrifugal forces. The highlight of findings are below.

It is evident that as particle inertia increases, a larger number of particles tend to accumulate in the outer wall region, leading to an increase in preferential concentration near the outer wall. To quantify the extent of particle accumulation near the outer wall, we employed two-dimensional Vorono{\"{i}} diagram analysis. While particles with low inertia exhibit a distribution close to random, an increase in particle inertia causes particles to cluster in the low-speed streaks at the outer wall, resulting in the formation of more clusters and voids. We further elucidated the emergence of strip-like structures by showing the ratios of azimuthal length to axial length for the Vorono{\"{i}} cells. Additionally, by counting the number of particles in regions of low and high azimuthal velocity, we observed a preference for particles to accumulate in low-speed fluid structures.

Then, we differentiated the effects of biased sampling, turbophoresis, and centrifugal forces by separately considering their relative strengths. By comparing the relative strength of these forces, we summarized the mechanisms of particle transport in Taylor-Couette flow in a schematic diagram depicted in figure~\ref{fig:11}. In terms of particle transport in the radial direction, centrifugal effects are predominant, with biased sampling effects being the second most influential for small inertial particles, while turbophoresis effects become second most important for large inertial particles. Each of these effects leads to distinct processes: centrifugal forces cause particles to migrate toward the outer wall; biased sampling forces cause particles to move toward the inner wall due to the fact that particles accumulate in the Taylor rolls with inward radial velocity; turbophoresis transports particles from the bulk region to the two walls due to the high turbulence intensity in the bulk region compared to that near the walls.

Finally, there are several crucial considerations that were not addressed in this study, which merit further investigation. For instance, at higher particle mass loading, turbulence is notably influenced by particle feedback forces, thus affecting both the turbulent flow and the particle distribution. The findings in this work provide groundwork for future study on the impact of inertial particles on particle distribution in rotating wall-bounded turbulent flows.

\section*{Acknowledgements}
This work was supported by the Natural Science Foundation of China under grant nos. 11988102, 92052201 and 12372219.

\section*{Declaration of interests}
The authors report no conflict of interest.

\appendix
\section{\label{Appendix}Derivation of the radial particle concentration profile} 
The wall-normal particle concentration profile in turbulent channel flow has been discussed by \citet{johnson2020turbophoresis}. Here we extend their proposed model to the radial particle concentration in Taylor-Couette flow with cylindrical coordinates. The derivation is based on the single-particle position-velocity probability density function (PDF) along radial direction. An additional centrifugal term emerges to account the effect of rotation.

The radial direction single-particle position-velocity PDF is denoted as $f(r, v_r, ; t)$, 
\begin{gather}
  f\left(r, v_r ; t\right)=\left\langle\delta(r-\hat{r}(t)) \delta\left(v_r-\hat{v}_r(t)\right)\right\rangle,
  \label{eq:A1}
 \end{gather}
where $\delta(x)$ is the Dirac delta function, ⟨·⟩ is the ensemble averaging over a disperse phase. Differentiating (\ref{eq:A1}) in time and substituting the particle radial dynamic equation: $\dot{\hat{r}}=\hat{v}_r; \dot{\hat{v}}_r=\hat{a}_r$, we have the evolution of $f(r, v_r, ; t)$:
\begin{gather}
  \frac{\partial f}{\partial t}+\frac{\partial\left(v_r f\right)}{\partial r}+\frac{\partial\left(\left\langle a_r | r, v_r \right\rangle f\right)}{\partial v_r}=\dot{f}_{\text {coll }},
  \label{eq:A2}
 \end{gather}
where the right hand side of (\ref{eq:A2}) is $\dot{f}_{\text {coll}}$ the particle collision term.

Since the particle phase in the turbulent Taylor-Couette flow is statistically homogeneous in the azimuthal and axial directions, The normalized particle concentration $C^*(r; t)$ can be obtained by integrating $f$ over the particle velocity $v_r$:
\begin{gather}
  C^*(r;t) = C_0 \int_{-\infty }^{\infty} f(r, v_r ; t)\mathrm{d}v_r. 
  \label{eq:A3}
 \end{gather}
Here the $C_0$ is the bulk particle concentration. The particle radial direction momentum conservation is obtained as a first-order moment of $f$, i.e. multiplying (\ref{eq:A2}) by $v_r$ and $C_0$ and integrating over particle radial velocity $v_r$:
\begin{gather}
  \frac{\partial\left(\left\langle v_r | r\right\rangle C^*\right)}{\partial t}+\frac{\partial\left(\left\langle v_r^2 | r\right\rangle C^*\right)}{\partial r}-\left\langle a_r |r\right\rangle C^*=0.
  \label{eq:A4}
\end{gather}
The particle collisional term can be neglected because of particle momentum conservation by each collision events. At statistically homogeneous state, the first term ${\partial\left(\left\langle v_r | r\right\rangle C^*\right)}/{\partial t}=0$, and the particle radial momentum balance equation can be simplified as:
\begin{gather}
  \frac{\mathrm{d}}{\mathrm{d} r}\left(\left\langle v_r^2 | r\right\rangle C^*\right)=\left\langle a_r | r\right\rangle C^*.
  \label{eq:A5}
\end{gather}
Substituting the particle radial acceleration $ a_r = (u_r - v_r)/St +{v_\theta^2}/{r}$ to the expression (\ref{eq:A5}), the particle radial momentum balance equation at statistically homogeneous state can be rewritten as:
\begin{gather}
  \left\langle v_r^2 | r\right\rangle \frac{\mathrm{d} C^*}{\mathrm{~d} r}=\left(\frac{\left\langle u_r | r\right\rangle}{St}-\frac{\mathrm{d}\left\langle v_r^2 | r\right\rangle}{\mathrm{d} r} + \frac{\left\langle v_\theta^2|r\right\rangle}{r} \right) C^*.
  \label{eq:A6}
\end{gather}
Then equation (\ref{eq:3.1}) can be obtained in the formal solution of (\ref{eq:A6}).

\bibliographystyle{jfm}
\bibliography{literatur}

\end{document}